# Fragmentation of expanding shells in spiral and irregular galaxies

**S. Ehlerová[1], J. Palouš[1], Ch. Theis[2], and G. Hensler[2]**

[1] Astronomical Institute, Academy of Sciences of the Czech Republic, Boční II 1401, 141 31 Prague 4, Czech Republic
[2] Institut für Astronomie und Astrophysik der Universität, D-24098 Kiel, Germany

**Abstract.** Conditions for the instability and the fragmentation of expanding and decelerating shells are discussed. A self-similar analytical solution is compared with the results of 3-dimensional computer simulations of expansions in homogeneous media. The amount of energy supply from the final number of young stars in an OB association, the value of the sound speed, the stratification and density of the ambient medium, the galactic differential rotation and the gravitational force perpendicular to the galactic plane influence the formation of fragments. The typical size of unstable shells is 1 kpc for density $n = 1$ cm$^{-3}$. In thick disk galaxies the fragmentation occurs in the nearly whole shell while in thin disks it is restricted to the galactic equator. Unstable fragments may become molecular and trigger the formation of molecular clouds where new stars are formed. We conclude that in dwarf galaxies the star formation may propagate in all directions turning the system into a starburst, contrary to spiral galaxies where the star formation propagates only in some directions in the thin strip near the symmetry plane.

## 1. Introduction

Radiation, ionization and mechanical energy input into the interstellar medium due to the evolution of massive stars and their supernova explosions are processes accelerating the ambient interstellar medium, agglomerating the gas into expanding and decelerating shells. The shells may become unstable, transforming themselves into molecular clouds which seed further star formation.

The shells expand into the interstellar medium with velocities exceeding the sound speed. The gravitational instability of shock-compressed layers has been investigated by Elmegreen & Elmegreen (1978), and a new type of hydrodynamic instability of decelerating pressure-driven shocks has been discussed by Vishniac (1983). The evolution of decelerating shock waves in a uniform density medium, or in a medium with large-scale density gradients, has been investigated in numerical studies (Mac Low & McCray, 1988; Mac Low et al., 1989; Yoshida & Habe, 1992). In low-density regions, where the hot gas accelerates the swept-up shell, the Rayleigh-Taylor instability develops. High-density regions are more sensitive to cooling, forming a radiative thin shell. Vishniac (1994) and Elmegreen (1989, 1994) discuss the combination of the hydrodynamical instability of radiative, expanding and decelerating shocks with the effects of self-gravity.

The linear analysis of this 'Elmegreen-Vishniac' instability is described in the next paragraph. Later, a solution similar to Sedov's (1959) solution of an expanding blast wave is adopted and compared with a three-dimensional computer model. Then, a more realistic distribution of ambient medium and the galactic differential rotation are included. Finally, the conditions for the shell fragmentation and star formation in spiral and irregular galaxies are discussed. The differences in the shell fragmentation may be the reason for changing conditions for star formation, resulting in different gas consumption rates between spiral and irregural galaxies.

## 2. Conditions for fragmentation and molecularization

The analysis of linear perturbations of transverse motions on a three-dimensional shell expanding into a uniform medium has been performed by Elmegreen (1994) and Vishniac (1994). Taking into account the convergence of the perturbed flow, stretching of the perturbed region due to expansion, and its own gravity, the instantaneous maximum growth rate $\omega$ of a transverse perturbation of a shell is given as

$$\omega = -\frac{3v}{R} + \sqrt{\frac{v^2}{R^2} + \left(\frac{\pi G \Sigma}{c}\right)^2}, \tag{1}$$

where $R$ is the radius of the shell with mass column density $\Sigma$. $v$ is its velocity of expansion relative to the ambient medium, $c$ is the sound speed within the shell, and $G$ is the constant of gravity. The instability occurs for

$$\omega > 0. \tag{2}$$

This is equivalent to

$$\xi \equiv \frac{\sqrt{8}vc}{\pi GR\Sigma} < 1, \tag{3}$$

where $\xi$ is a dimensionless parameter. This maximum growth rate corresponds to the transversal angular wavenumber

$$\eta = \frac{\pi G\Sigma R}{c^2}. \tag{4}$$

Another condition of the instability is that the wavelength of the fastest transversal perturbation $\lambda$ is within a fraction of the shell, or smaller than $R$, which may be written as

$$\lambda = \frac{2\pi}{\eta} R < R, \tag{5}$$

which is equivalent to

$$\frac{2c^2}{G\Sigma R} < 1. \tag{6}$$

This may be rewritten as

$$\frac{\pi\xi}{\sqrt{2}M} < 1 \tag{7}$$

where $M \equiv v/c$. This is only a restriction to the $\xi$ parameter if $M$ is near unity. For $M \gg 1$, which is satisfied in almost all the cases examined, the minimum wavelength of a growing transverse perturbation is always within a fraction of the shell radius.

The fragmentation begins when both criteria are fulfilled for the first time, $t = t_b$. We follow the expansion even beyond this time and evaluate the fragmentation integral $I_f(t)$:

$$I_f(t) = \int_{t_b}^{t} \omega(t')dt'. \tag{8}$$

At the time $t = t_f$ when $I_f(t) = 1$ the fragments are well developed, so that clouds form at a later time.

In the following sections, the models of an expanding shell are described, and the above instability criteria are tested. As soon as $t = t_f$, the approximation of an expanding shell probably breaks down, and the fragments continue along individual galactic orbits.

If the particle column density $N$ in a fragment surpasses the critical value (Franco & Cox, 1986)

$$N_{crit} = 5 \times 10^{20} \frac{Z_\odot}{Z} cm^{-2}, \tag{9}$$

(where $Z$ is the metallicity and $Z_\odot$ is the solar value) the gas is shielded against the ionizing interstellar radiation, can recombine and later become molecular. If this condition is fulfilled during $t_b < t < t_f$, we conclude that the expansion of the shell triggered the formation of a new molecular cloud.

## 3. Self-similar solution

Using a dimensional analysis, Sedov (1959) derived a solution for supersonic, spherical expansion of a shell into the homogeneous medium in case of a single instantaneous energy input and vanishing external pressure. For the case of interstellar wind bubbles Castor et al. (1975) and Weaver et al. (1977) derived the solution which describes the structure and evolution of the shocked stellar wind region (their zone b). This case is characterized by steady energy and mass input rates. Since this should be more appropriate for the temporal energy injection by type II supernovae events in an OB association, we applied the same method here. The solution gives the relation between the radius of the shell $R$, the supernova rate $\dot{N}_{SN}$, the mass density of the ambient medium $\rho_0$, and the expansion time $t$ as follows ($E_{SN}$ is the energy release per supernova):

$$R(t) = \left(\frac{125}{154\pi}\right)^{1/5} \left(\frac{\dot{N}_{SN} \cdot E_{SN}}{\rho_0}\right)^{1/5} t^{3/5}$$

$$= 53.1 \cdot \left[\left(\frac{\dot{N}_{SN}}{Myr^{-1}}\right)\left(\frac{E_{SN}}{10^{51}erg}\right)\right]^{1/5}$$

$$\left[\left(\frac{\mu}{1.3}\right)^{-1}\left(\frac{n_0}{cm^{-3}}\right)^{-1}\right]^{1/5} \cdot \left(\frac{t}{Myr}\right)^{3/5} pc, \tag{10}$$

where $\mu$ is the average molecular weight and $n_0$ is the number density of particles in the ambient medium. The expansion velocity of the shell $v$ is given by the time derivative of Eq. (10). Assuming that the total mass $m(R) = 4\pi/3\,\rho_0 R^3(t)$ is swept up in the shell, the surface density is given by

$$\Sigma(t) = \frac{m(R)}{4\pi R^2(t)} = \left(\frac{125}{37\,422\pi}\right)^{1/5} \left(\dot{N}_{SN} \cdot E_{SN} \cdot \rho_0^4\right)^{1/5} t^{3/5}$$

$$= 0.57 \left[\left(\frac{\dot{N}_{SN}}{Myr^{-1}}\right)\left(\frac{E_{SN}}{10^{51}erg}\right)\right]^{1/5}$$

$$\left[\left(\frac{\mu}{1.3}\right)^{4}\left(\frac{n_0}{cm^{-3}}\right)^{4}\right]^{1/5} \cdot \left(\frac{t}{Myr}\right)^{3/5} M_\odot/pc^2 \tag{11}$$

Inserting eqs. (10) and (11) in eq. (3) we get the instability parameter $\xi$:

$$\xi(t) = \frac{24\sqrt{2}}{5} \left(\frac{18711}{64000\,\pi^4}\right)^{1/5} \cdot$$

$$\cdot \frac{c}{G} \left(\frac{1}{\rho_0^4 \dot{N}_{SN} \cdot E_{SN}}\right)^{1/5} t^{-8/5} \tag{12}$$

The temporal evolution of $\xi(t)$ defines an 'instability time' $t_b$, i.e. the onset of gravitational instability, which can be evaluated by the condition $\xi(t_b) = 1$, or

$$t_b = 28.8 \left(\frac{c}{1km/s}\right)^{5/8} \left(\frac{n_0}{cm^{-3}}\right)^{-1/2} \left(\frac{\mu}{1.3}\right)^{-1/2} \cdot$$

$$\left(\frac{E_{SN}}{10^{51}erg}\right)^{-1/8} \left(\frac{\dot{N}_{SN}}{Myr^{-1}}\right)^{-1/8} Myr \tag{13}$$

Eq. (12) shows that the system is always gravitationally stable prior to $t_b$ and becomes unstable at later times. The radius of the shell at $t_b$ is

$$R(t_b) = 399 \left(\frac{c}{1\mathrm{km/s}}\right)^{3/8} \left(\frac{n_0}{\mathrm{cm}^{-3}}\right)^{-1/2} \left(\frac{\mu}{1.3}\right)^{-1/2} \cdot \left(\frac{E_{\mathrm{SN}}}{10^{51}\mathrm{erg}}\right)^{1/8} \left(\frac{\dot{N}_{\mathrm{SN}}}{\mathrm{Myr}^{-1}}\right)^{1/8} \mathrm{pc},$$ (14)

and the expansion velocity

$$v(t_b) = 8.13 \left(\frac{c}{1\mathrm{km/s}}\right)^{-1/4} \cdot \left(\frac{E_{\mathrm{SN}}}{10^{51}\mathrm{erg}}\right)^{1/4} \left(\frac{\dot{N}_{\mathrm{SN}}}{\mathrm{Myr}^{-1}}\right)^{1/4} \mathrm{km/s}.$$ (15)

The expansion velocity at the fragmentation time is independent of the density of the ambient medium. This may be one of the reasons why the random motion in the interstellar medium is almost constant over a broad range of densities. $t_b$, $R(t_b)$ and $v(t_b)$, according to formulae (13), (14) and (15) are shown together with the results of 3D computer simulations (see below) in Fig. 1.

In order to calculate the 'fragmentation time' $t_f$, i.e. the time when the fragmentation becomes strongly nonlinear, we can rewrite the growth rate (1) as

$$\omega = -\frac{v}{R} \left(-3 + \sqrt{1 + \frac{8}{\xi^2(t)}}\right)$$ (16)

and apply the relation $\xi(t) = (t/t_b)^{-8/5}$ together with the Sedov solution (10). From the condition $\int_{t_b}^{t_f} \omega(t)\, dt = 1$, we get

$$\frac{5}{3} = \int_1^{x_f} \frac{-3 + \sqrt{1 + 8x^{16/5}}}{x}\, dx$$ (17)

with $x_f \equiv t_f/t_b$. It should be noted that the ratio of the fragmentation time $t_f$ to the instability time $t_b$ is independent of the density of the ambient medium and of the energy injection rate and the sound speed in the shell. For a homogeneous medium a numerical solution of Eq. (17) gives $x_f \approx 2.03$. This value will be compared with the results of 3D computer simulations below.

The timescales themselves are also almost independent on the energy injection rate, but depend on the density of the ambient medium. For a typical density in the solar neighbourhood of $n = 1\ cm^{-3}$, the fragmentation time is of the order of $(5-6) \cdot 10^7$ yr. The corresponding radii of the shells are shown in Fig. 2: for $n = 1\ cm^{-3}$ we obtain radii of 600 pc to 1.1 kpc for sound speeds within the shell $c$ between $1\ km/s$ and $5\ km/s$, respectively. In Fig. 3 we give the $\dot{N}_{\mathrm{SN}}$ as a function of the ambient medium density $n_0$, which is necessary to reach the critical value of column density $N_{crit}$ (9) at times $t_b$ and $t_f$ required for molecularization. This diagram gives the lower limit for $\dot{N}_{\mathrm{SN}}$ at given $n_0$ for the formation of molecular clouds from fragmenting shells.

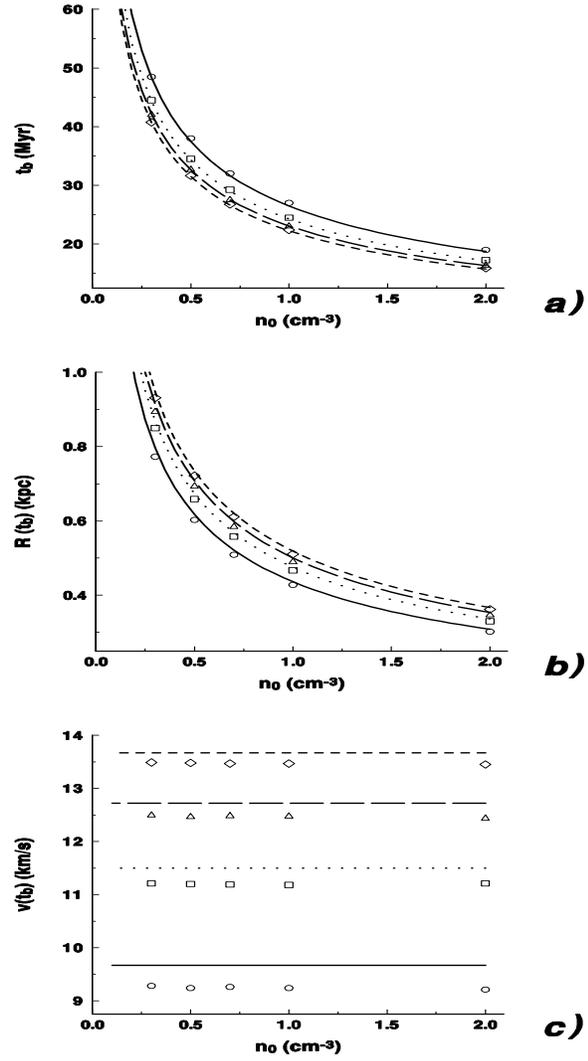

**Fig. 1a–c.** The instability time $t_b$ (**a**), the shell radius $R(t_b)$ (**b**) and the expansion velocity $v(t_b)$ (**c**) as a function of $n_0$ for different $\dot{N}_{\mathrm{SN}}$. Continuous lines are the solution as given by formulae (13), (14) and (15), symbols show results of computer simulations. ○ corresponds to $\dot{N}_{\mathrm{SN}} = 2\ Myr^{-1}$, □ to $\dot{N}_{\mathrm{SN}} = 4\ Myr^{-1}$, △ to $\dot{N}_{\mathrm{SN}} = 6\ Myr^{-1}$, and ◇ to $\dot{N}_{\mathrm{SN}} = 8\ Myr^{-1}$.

Since gravitational instability is a condition for the onset of star formation, radii from Fig. 2 and $\dot{N}_{\mathrm{SN}}$ from Fig. 3 give a lower limit for distances between the subsequent star-forming regions and a lower limit for the required $\dot{N}_{\mathrm{SN}}$ in an OB association to propagate star formation at a given density of the ambient medium $n_0$. If the propagating star formation is induced by instabilities in the shells, we may also estimate the speed of propagation.

If the OB stars are exploding inside a molecular cloud of $n \approx 10^3 - 10^4 cm^{-3}$, radii at the fragmentation time shrink to 10–20 pc which are smaller than typical tidal radii for a $10^5$–$10^6\ M_\odot$ stellar system in a $10^{11}\ M_\odot$ galaxy. Therefore, at high densities in the ambient medium the system of fragments can form quickly enough in order to prevent a tidal disruption of the shell. Finally, these clumps might recollapse and form a gravitationally bound system as suggested by Brown et al. (1991).

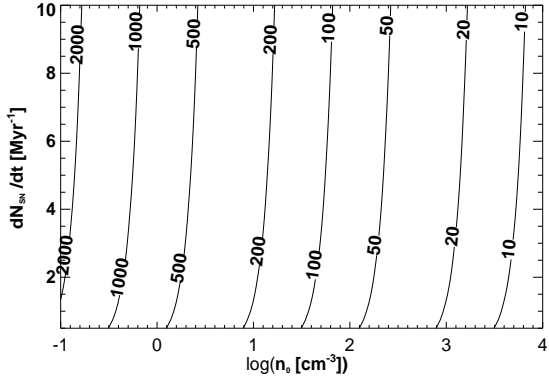

**Fig. 2.** The contour lines show the radius of the shell (in pc) at the fragmentation time $t_f$ as a function of the supernova rate $\dot{N}_{SN}$ and the number density $n$ in a homogeneous ambient ISM. The sound speed inside the shell was assumed to be $c = 1$ km/s and the molecular weight was set to $\mu = 1.3$.

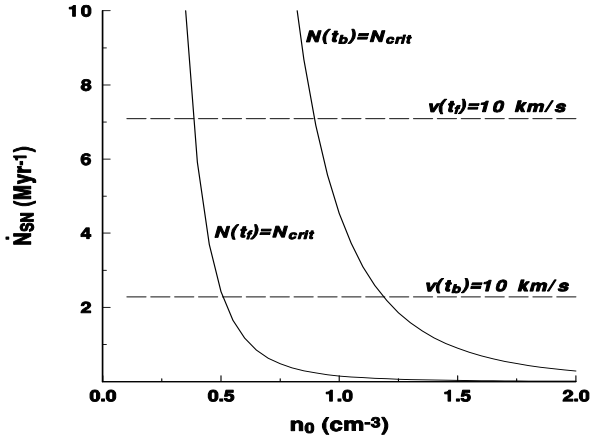

**Fig. 3.** Solid lines show the value of $\dot{N}_{SN}$ required by condition (9) at $t_b$ and at $t_f$ as a function of the density $n_0$ of the ambient medium. Dashed lines indicate constraints on the $\dot{N}_{SN}$ implied by the non-zero sound speed in the ambient medium (see Sect. 6).

## 4. 3D numerical simulations

The expansion of gas layers around OB associations may be described as a blastwave propagating into the interstellar medium (Ostriker & McKee, 1988; Bisnovatyi-Kogan & Silich, 1995). A multi-supernova remnant follows an initial quasi-adiabatic stage, driven by the thermalized energy of the supernova ejecta. The energy from each new supernova is released within the remnant created by all previous stellar winds and explosions. Once the radiative losses become important a thin and cold shell forms at the outer edge of the cavity, which is filled by the hot medium.

Due to its supersonic speed the shell continues to collect the ambient medium and it slows down to velocities comparable to the random motion of the interstellar medium.

Since the radius of the shell $R$ is much larger than the thickness of it, the thin layer approximation considered by Sedov (1959) and developed by Kompaneets (1960) and Bisnovatyi-Kogan & Blinikov (1982) can be applied. This approximation has been adopted in two dimensional models of expanding shells by Tenorio-Tagle & Palouš (1987), Mac Low & McCray (1988) and Palouš et al. (1990) and compared with the 2D hydrodynamical simulations by Mac Low et al. (1989). These models have been further extended into three dimensions by Palouš (1990, 1992), Ehlerová & Palouš (1996) and by Silich et al. (1996).

Here, the 3D computer model of an expanding infinitesimally thin shell as described by Ehlerová & Palouš (1996) is modified. Later, the results are compared with the Sedov solution.

### 4.1. Initial conditions

At an initial time $t_{ini}$ we insert an initial energy $E_{ini}$ into a spherical cavity of a small initial radius $R_{ini}$. The values of $R_{ini}$, $E_{ini} = \dot{N}_{SN} \times E_{SN} \times t_{ini}$, and $t_{ini}$ have to fulfill together with the given density of the ambient medium $\rho_0$ the relation given by Eq. (10). In Sedov's solution 19% of the initial SN energy is transformed into kinetic energy of the expanding shell $E_{sh}$ and the rest into thermal energy of the hot bubble $E_{int}$:

$$E_{ini} = E_{int} + E_{sh}. \tag{18}$$

The internal pressure $P_{int}$ can be computed as

$$P_{int} = \frac{2}{3} \frac{E_{int}}{V_{ini}}, \tag{19}$$

where $V_{ini}$ is the initial volume corresponding to a sphere with the radius $R_{ini} \sim 10$ pc. The initial mass in this spherical shell is the total mass inside the initial volume. This defines together with the initial kinetic energy $E_{sh}$ the initial expansion velocity of the shell $v_{ini}$.

### 4.2. Energy inputs from sequential supernovae

During their evolution, SN explosions supply energy to the bubble. The rate of energy input is given by $\dot{N}_{SN}$, which we define in units of supernova events per $10^6$ years. We always take as the energy released by a single supernova a value $E_{SN} = 10^{51}$ erg. We also give the total number of supernovae which may deliver the energy from an OB association. When this number is reached the energy input stops.

In previous simulations (Ehlerová & Palouš, 1996) with the abrupt injection of the energy, the pressure was computed using the adiabatic equation ($PV^\gamma = const$). The temperature was then determined by the equation of state. There was no apparent connection between pressure and thermal energy.

In this paper, the internal thermal energy is evaluated at every timestep, and subsequently the pressure and temperature are derived. The internal energy of a bubble changes due to:

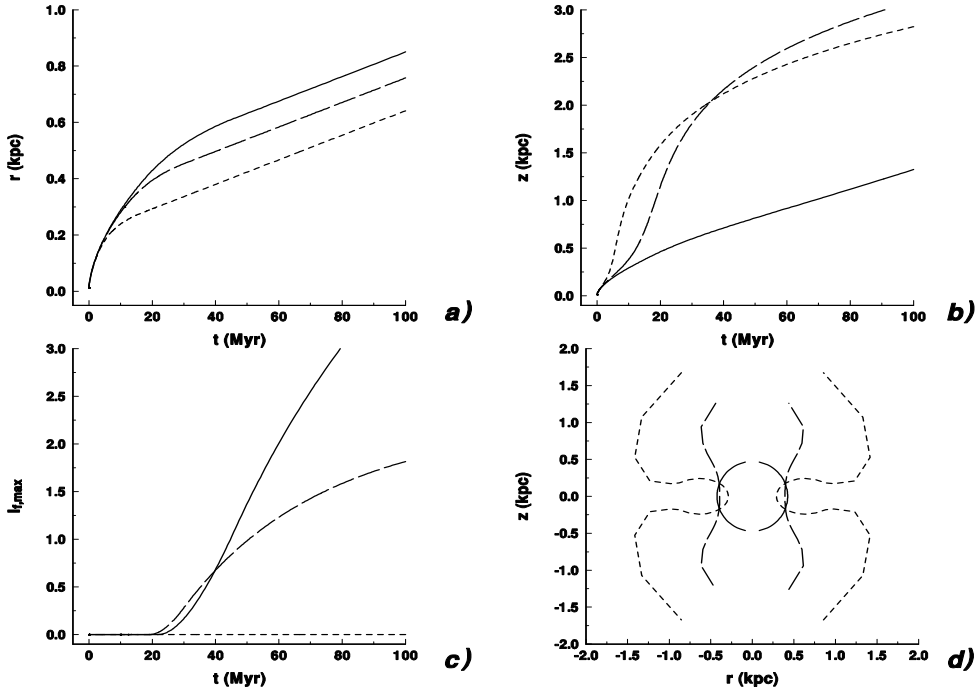

**Fig. 4a–d.** Radius $r$ (**a**), vertical extension $z$ (**b**) and the fragmentation integral $I_{f,max}$ (**c**) as a function of time, and the shape of the shells after 20 Myr of expansion (**d**) for different values of $\sigma_z$. Solid line: $\sigma_z = 500$ pc, long dash line: $\sigma_z = 200$ pc, short dash line $\sigma_z = 100$ pc.

– SN explosions; $\Delta E_1 = E_{SN} \times \dot{N}_{SN} \times \Delta t$,
– work done by the pressure; $\Delta E_2 = P \, \Delta V$,

where $\Delta V$ is the increment of the bubble volume over the timestep $\Delta t$.

We have

$$E_{int,new} = E_{int,old} + \Delta E_1 - \Delta E_2, \qquad (20)$$

where $E_{int,old}$ and $E_{int,new}$ are the values of the internal energy before and after one step in time.

### 4.3. Integration scheme

In the 3D numerical simulations, the shell is divided into $N_1$ layers and every layer into $N_2$ elements. The motion of each of the $N_1 \times N_2$ elements is solved numerically. The momentum conservation equation is given as

$$\frac{d}{dt}(mv) = dS[(P_{int} - P_{ext}) + \rho_0 v_0(v - v_0)] + mg, \qquad (21)$$

where $m$, $v$, $dS$ are the mass, expansion velocity and the surface of an element, respectively, $P_{int}$ and $P_{ext}$ are the pressures inside and outside of the bubble, $\rho_0$ and $v_0$ are the density and velocity of the ambient medium, and $g$ is the gravitational acceleration.

The mass conservation equation gives the increase of mass $m$ as long as the expansion velocity component normal to the shell, $v_\perp$, exceeds the velocity of sound in the ambient medium:

$$\frac{d}{dt}m = v_\perp \rho_0 dS. \qquad (22)$$

After the expansion becomes subsonic, the mass $m$ of an element is constant.

The momentum conservation equation (21) and mass conservation equation (22) are solved numerically together with the equations for the internal energy (20) using finite timesteps. An adaptive step-size control scheme is used, which will be described elsewhere. The main advantages of this scheme are the known accuracy of the integrated quantities and the savings of the CPU time, particularly in the subsonic stage where the timestep can be large without loosing the accuracy.

## 5. Comparison of 3D numerical simulations with Sedov solution

In order to evaluate the fragmentation criteria (3) and (7) to derive the instability time $t_b$, and $t_f$ from the fragmentation integral $I_f$ according to the equation (8), we have to estimate the value of the sound speed within the shell $c$ which we assume to be constant over a reasonable fraction of the expansion. As long as the cold and thin shell keeps its low temperature, the reverse shock increases the thickness of the shell keeping the volume density inside the shell constant, which results in a constant sound speed. We take arbitrarily the value $c = 1$ km/s, which is probably the lower (and therefore the most suitable) limit.

The results from the computer simulations are compared to formulae (13), (14) and (15) from the Sedov solution in Fig. 1. There is very close agreement between all the predicted and computed values. The 3D simulations confirm that $t_b$ and $R(t_b)$ are proportional to $n_0^{-1/2}$ and that $v(t_b)$ is fairly independent of it. The slightly lower expansion velocities at $t_b$ from simulations reflect the fact that the Sedov solution omits the pressure of the ambient medium and that it does not take into account the work done by bubble expansion.

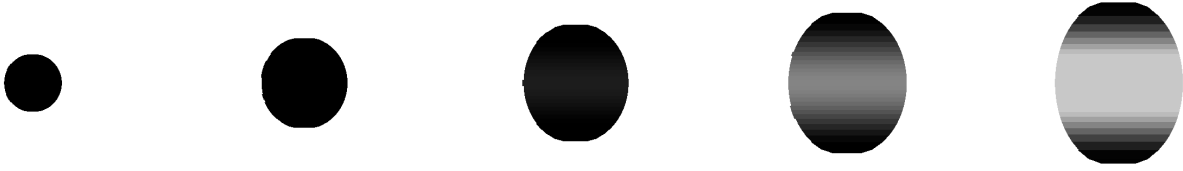

**Fig. 5.** The time sequences of shell expansion in the one component thick gaussian disk with $\sigma_z = 500$ pc. The snapshots are for $t = 10, 20, 30, 40$ and 50 Myr. A black color shows regions where the value of fragmentation integral $I_f = 0$, a white color shows regions where the value $I_f \geq 1$. The black line has a length of 1 kpc.

The fragmentation integrals and $t_f$ are also evaluated. We can conclude that the value of $t_f/t_b$ from simulations is always in the interval (2.01, 2.03), which is rather close to 2.03 derived from the solution of Eq. (17). Knowing $t_f$ we can also check the expansion velocity $v(t_f)$:

$$v(t_f) = 0.73 \, v(t_b). \tag{23}$$

The value of $v(t_f)$ is in all the cases examined less than 10 km/s: 6.5 km/s $< v(t_f) <$ 10 km/s, which is in very good correspondence with the observed values of random motions in the interstellar medium.

## 6. Exhaustion of the energy supplies and the sound speed in the ambient medium

The energy supply from SN into the bubble is limited by the finite number of massive stars in an OB association. In our Galaxy this number has been estimated by Tenorio-Tagle & Bodenheimer (1988) to 40 – 100. When the energy input stops, the supersonic expansion speed decelerates more rapidly than according to the self-similar solution (10), reaching the value of the sound speed in the ambient medium earlier at lower shell surface densities. This implies that the fragmentation conditions (3) and (7) may not be fulfilled. This is particularly important in the low density interstellar medium, where the deceleration of expanding shells proceeds slowly and the value of $t_b$ is large.

Another constraint is given by the value of the sound speed in the ambient medium $c_{amb}$. The accumulation of mass into the shell stops when $v$ decelerates to $c_{amb}$, which happens at time $t_{sound}$ (for the uninterrupted energy input given by the following formula):

$$
\begin{aligned}
t_{sound} &= \left(\frac{3}{5}\right)^{\frac{5}{2}} \left(\frac{125}{154\,\pi}\right)^{\frac{1}{2}} \left(\frac{\dot{N}_{SN} \times E_{SN}}{\rho_0}\right)^{\frac{1}{2}} (c_{amb})^{-\frac{5}{2}} \\
&= 97.3 \left(\frac{\dot{N}_{SN} \times E_{SN}}{10^{51} erg Myr^{-1}}\right)^{\frac{1}{2}} \left(\frac{n}{cm^{-3}}\right)^{-\frac{1}{2}} \\
&\quad \left(\frac{\mu}{1.3}\right)^{-\frac{1}{2}} \left(\frac{c_{amb}}{5\,km\,s^{-1}}\right)^{-\frac{5}{2}} Myr.
\end{aligned}
\tag{24}
$$

The shell motion continues as a sound wave. Such subsonic expansion dilutes the shell decreasing its surface density $\Sigma$ and increasing the value of the instability parameter $\xi$, which may imply that the fragmentation conditions (3) and (7) are again not fulfilled.

We distinguish three cases:

- for $t_{sound} < t_b$ the shell never fragments;
- for $t_b < t_{sound} < t_f$ the fragments form if at $t_{sound}$ the value of the fragmentation integral $I_f$ is close to 1. If $I_f(t_{sound})$ is close to 0 the fragments reexpand again;
- for $t_f < t_{sound}$ the fragments detach from the shell moving along individual ballistic orbits in galaxy.

It has been shown in the previous section that for spherical shells with continuous energy supply $v(t_f)$ is lower than 10 km/s. When the energy supply stops before $t_f$ the velocity $v(t_f)$ is even lower.

As it is shown in Fig. 3 for $c_{amb} = 10$ kms$^{-1}$ the fragments form at $t_b$ or $t_f$ when the values of $n_0$ and $\dot{N}_{SN}$ are to the right of the solid lines. They are molecular if $n_0$ and $\dot{N}_{SN}$ are above the dashed lines. For a homogeneous, non–stratified distribution of the ambient medium, and with OB associations composed of 40 – 100 massive stars, we can separate: $n_0 > 1$ cm$^{-3}$, which gives $t_f < t_{sound}$ and the shell fragments; $0.3 < n_0 < 1$ cm$^{-3}$, which implies $t_b < t_{sound} < t_f$, and the shell will fragment only in some cases; $n_0 < 0.3$ cm$^{-3}$, which yields $t_{sound} < t_b$, and the fragments will never form.

## 7. Stratified distribution of the interstellar medium

The distribution of the interstellar medium in the $z$ direction perpendicular to the galactic plane is approximated with one or more Gaussian and exponential components (Dickey & Lockman, 1990). Here, we examine the expansion of bubbles in a medium with $z$−stratification and discuss how the shell fragmentation is influenced.

For simplicity, we start with one Gaussian component

$$n(z) = n_0 \times \exp\left(-\frac{z^2}{\sigma_z^2}\right), \tag{25}$$

where $\sigma_z$ is the gaussian half-thickness of the disk. Later, the discussion will be extended also to multi-component disks.

The results of computer simulations are presented in Fig. 4, where the radius of the shell in the galactic plane $r$, the maximum distance of the shell to the galactic plane $z$ and the maximum value of the fragmentation integral $I_{f,max}$, are given as functions of time for different values of $\sigma_z$.

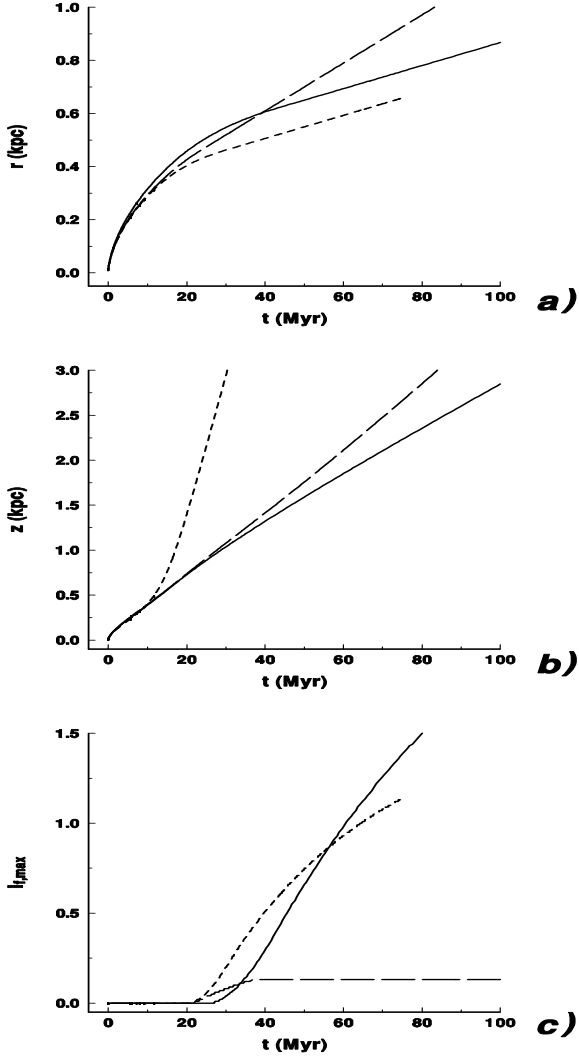

**Fig. 6a–c.** The value of $r$, $z$ and $I_{f,max}$ are given as functions of time. Solid line: multi-component disk; long dash line: one component exponential disk; short dash line: one component gaussian disk.

The time sequence of snapshots for the case of the thick disks, with $\sigma_z \sim 500$ pc is shown in Fig. 5. The evolution is similar to the case of a homogeneous medium. Due to the density gradient perpendicular to the plane the shell is slightly stretched in the $z$-direction, which is shown with larger values of $z$ in relation to $r$ for given time. The instability belt, where the fragmentation conditions (3) and (7) are fulfilled grows with time, and before the time $t_f$ in the most unstable parts near the galactic plane is reached, almost all the shell becomes unstable.

In disks of medium thickness, $\sigma_z \sim 200$ pc, the shells are rather elongated reaching $z \sim 1$ kpc, and the instability belt is restricted to $z < 100$ pc only. Parts of the shell at larger heights are always gravitationally stable.

In thin disks, $\sigma_z \lesssim 100$ pc the bubble blows out to $z \sim 3$–4 kpc. It is always stable, even near to the galactic plane. There the instability conditions are fulfilled at some time, however the shell decelerates to the speed of sound well before $t_f$ and the fragments reexpand (see Fig. 4).

For a comparison, a multi-component disk as proposed for the solar vicinity in the Milky Way by Dickey & Lockman (1990) was tested. The $z$ distribution of interstellar medium is given as

$$n(z) = 0.395 \times \exp\left(-\frac{z^2}{\sigma_1^2}\right) + 0.107 \times \exp\left(-\frac{z^2}{\sigma_2^2}\right) +$$
$$+ 0.064 \times \exp\left(-\frac{|z|}{h_z}\right) cm^{-3}, \tag{26}$$

where $\sigma_1 = 212$ pc, $\sigma_2 = 530$ pc and $h_z = 403$ pc. This multi-component disk is compared to a disk with only one gaussian or one exponential component of the scale-heights 200 pc. All the three disks have the same surface density 9.6 $M_\odot pc^{-2}$. The results of simulations are shown in Fig. 6. In the multi-component disk the shells reach smaller distances from the galactic plane than in disks with one gaussian or exponential component, and the fragmentation integrals evolve with time in a similar way as in the case of the disk with one gaussian component.

## 8. Galactic rotation and $K_z$ force

Flat disk galaxies rotate rapidly and the rotation influences the shapes of expanding shells: in the galactic plane, shells are not spherical any more but elliptical. In agreement with this result, many HI holes discovered in nearby galaxies (Brinks, 1994) have elliptical shapes. Due to galactic shear, the mass accumulated in the shell slides to the tips. This changes the fragmentation conditions in the galactic plane forming instability regions around the tips of elongated shells near the symmetry plane of the galaxy.

The plane parallel component of the gravitational acceleration is derived from the galactic rotation curve. We adopt the rotation curve proposed for the Milky Way by Wouterloot et al. (1990)

$$V(R_{gal}) = V_0 \left(\frac{R_{gal}}{R_0}\right)^{0.0382}, \tag{27}$$

where $R_{gal}$ is the galactocentric distance, $R_0 = 8.5$ kpc, $V(R_{gal})$ is the linear velocity of rotation and $V_0 = 220$ km/s.

The $z$-component of the galactic gravitational acceleration is approximated by a formula (Kuijken & Gilmore, 1989)

$$g_z = -2\pi G z \left[\frac{\Sigma_D}{(z^2 + z_D^2)^{1/2}} + 2\rho_h\right], \tag{28}$$

where $\rho_h = 0.015$ $M_\odot pc^{-3}$ is the halo density, $\Sigma_D = 46$ $M_\odot pc^{-2}$ and $z_D = 0.3$ $kpc$ are the surface density and scale height of the stellar disk near the Sun.

The results of our 3D computer simulations with the galactic rotation given by formula (27) and the component of the force perpendicular to the galactic plane given by formula (28) in the multi-component disk are shown as the time sequence of snapshots in Fig. 7. The instability develops only in the regions near the elliptical tips at $z$ close to 0. All the rest of the expanding shell is stable.

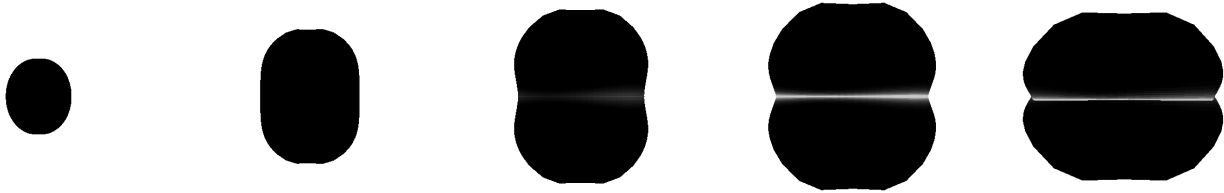

**Fig. 7.** The time sequences of shell expansion including the galactic differential rotation, component of force perpendicular to the galactic plane in the multi-component disk. The snapshots are for $t = 10, 20, 30, 40$ and 50 Myr, a black color shows regions where the value of fragmentation integral $I_f = 0$, a white color shows regions where the value $I_f \geq 1$. The black line has a length of 1 kpc.

## 9. Shells in spiral and dwarf galaxies

In this paper we have addressed the basic question of the conditions for triggered star formation in superbubble shells. Although physically reasonable and therefore expected to occur in general, only a few examples are observed where star formation sites are located at the rim of bubbles and just embedded into shells. As the closest case to us our local bubble could be envoked where Lindblad's ring seems to house the Orion, Perseus, and Sco-Cen OB associations on its periphery (Elmegreen, 1992). Comerón et al. (1993) and Comerón & Torra (1994) show that Cygnus OB1 - OB9 may be due to gravitationally unstable perturbations of the Cygnus Superbubble, which is powered by the Cygnus OB2 association. As examples of more distant star-forming loops which are connected with shells resulting from supernovae or accumulated stellar winds, Schwarz (1987) has presented e.g. the Mon OB 1 loop, Cep OB 3 loop, and a few others. In addition, G54.4-0.3 (Junkes et al., 1992a, 1992b) reveals the striking features of triggered star formation connected with a supernova shell.

Because observations of star-forming supernova shells within our Milky Way are difficult to perform in complementary spectral ranges since they are affected e.g. in X-rays by interstellar hydrogen absorption and in the UV and optical by dust extiction, extragalactic candidates could serve for better studies, e.g. in the LMC. The LMC-2 bubble (Wang & Helfand, 1991), LMC-4 bubble (Vallenarí et al., 1993), Constellation III (Dopita et al., 1985) and also 30 Dor reveal impressive structures of present star-forming episodes triggered and located in giant shells. (We refer the interested reader to the comprehensive paper on triggered star formation by Elmegreen, 1992).

The holes in the HI distribution of spiral galaxies like M 31 (Brinks & Bajaja, 1986) or in dwarf galaxies like Ho II (Puche et al., 1992) are in most cases due to the energy released from young stars and we may ask the question if the expanding shells can fragment and possibly trigger the next generation of the star formation: The observations may be compared with computer simulations of the shell expansion, so that we can discuss whether the conditions for fragmentation and molecularization can be fulfilled.

The evolution of an expanding shell in a dwarf galaxy with the thick HI disk is shown in Fig. 5. It fragments almost everywhere. The fragments become molecular if the density of the ambient medium is high enough according to Fig. 3. Thus the star formation in dwarf galaxies propagates in all directions, with the speed of propagation of $20 - 30$ km s$^{-1}$ turning all the galaxy into a starburst within a few $\times 10^8$ yr.

The evolution of a shell in a spiral galaxy similar to Milky Way composed of multi-component disk where the galactic differential rotation and the force perpendicular to the galactic plane are included is given in Fig. 7. The fragmentation happens close to the galactic plane in an unstable region near the tips of the elongated bubble. In spiral galaxies, the star formation can propagate only in some directions in the galaxy plane with a lower speed than in dwarf galaxies of $\sim 15$ km s$^{-1}$. The propagating star formation may result in spiral structures (Jungwiert & Palouš, 1994), however it can never turn a spiral galaxy into a starburst.

## 10. Conclusions

Our results provide a first insight into the required conditions for the development of Elmegreen-Vishniac instabilities within the expanding and decelerating shells giving constraints on the propagation of star formation. The density gradients, the $K_z$ force and the galactic shear distorting the expanding and decelerating shells from spherical symmetry are considered in computer simulations. These effects can enhance the fragmentation in the galactic plane and at the tips of elliptical shells, and supress it at other places. The newly formed clouds, possibly gravitationally stable at the time of their formation, may become unstable at later times and transform themselves into new places of star formation. Deviations from our results may be due to preexisting inhomogeneities, streaming motions connected with the spiral arms and small-scale density gradients.

Fragmentation of expanding shells may be one of the mechanisms how the star formation propagates in galaxies. In dwarf galaxies similar to LMC, SMC or Holmberg II, the conditions for fragmentation are fulfilled in almost any part of the expanding shell, therefore, the star formation can easily propagate. In the Milky Way-type spiral galaxies, where the fragmentation is restricted only to a rather narrow strip near the galactic plane, the star formation can propagate due to the fragmentation of expanding shells only into a small fraction of the interstellar space. At large heights above the galactic disk the fragmentation is almost prevented. In these galaxies the star formation is related to other large-scale mechanisms such as spiral density-waves, bars, galaxy mergers and interactions of galaxies.

*Acknowledgements.* The authors gratefully acknowledge financial support by the Grant Agency of the Czech Republic under grant no. 205/97/0699 and by the Deutsche Forschungsgemeinschaft under grant no. 436 TSE 17/5/96. The hospitality of the Institut für Astronomie und Astrophysik in Kiel and the Astronomical Institute of the Academy of Sciences of the Czech Republic in Prague is highly appreciated. We would like to thank the referee, E. T. Vishniac, for valuable remarks and comments and W. Walsh for careful reading of the text.